\documentclass[aps,prl,amsmath,amssymb,floatfix,twocolumn,amsmath,superscriptaddress,twocolumn,nofootinbib,tighten,letterpaper]{revtex4}
\usepackage{multirow}
\usepackage{bbold}
\usepackage{subfigure}
\usepackage{color}
\usepackage{mathrsfs}
\usepackage{hyperref}
\usepackage[normalem]{ulem}
\usepackage{bm}

\usepackage{amsfonts, relsize, color}
\usepackage{graphics}
\usepackage{graphicx}
\usepackage{subfigure}
\usepackage{hyperref}
\usepackage{color}

\begin{document}
\title{\bf Fermionic multicriticality near Kekul\'{e} valence-bond ordering in honeycomb lattice}

\author{Bitan Roy}\email{bitanroy@pks.mpg.de}
\affiliation{Max-Planck-Institut f\"{u}r Physik komplexer Systeme, N\"{o}thnitzer Str. 38, 01187 Dresden, Germany}

\author{Vladimir Juri\v ci\' c}\email{vladimir.juricic@nordita.org}
\affiliation{Nordita, KTH Royal Institute of Technology and Stockholm University, Roslagstullsbacken 23,  10691 Stockholm,  Sweden}

\date{\today}
\begin{abstract}
We analyze the possibility of emergent quantum multicritical points (MCPs) with enlarged chiral symmetry, when strongly interacting gapless Dirac fermions acquire comparable propensity toward the nucleation of Kekul\'{e} valence-bond solid (KVBS) and charge-density-wave ($N_b=1$) or $s$-wave pairing ($N_b=2$) or anti-ferromagnet ($N_b=3$) in honeycomb lattice, where $N_b$ counts the number of bosonic order parameter components. Besides the cubic terms present in the order parameter description of KVBS due to the breaking of a discrete $Z_3$ symmetry, quantum fluctuations generate \emph{new} cubic vertices near the high symmetry MCPs. All cubic terms are \emph{strongly} relevant at the bare level near three spatial dimensions, about which we perform a leading order renormalization group analysis of coupled Gross-Neveu-Yukawa field theory. We show that due to non-trivial Yukawa interactions among gapless bosonic and fermionic degrees of freedom, all cubic terms ultimately become \emph{irrelevant} at an $O(2+N_b)$ symmetric MCP, at least near two spatial dimensions, where $N_b=1,2,3$. Hence, MCPs with an enlarged $O(2+N_b)$ symmetry near KVBS ordering are \emph{stable}.       
\end{abstract}

\maketitle

\emph{Introduction}: Gapless Dirac fermions constitute an ideal arena to explore the effects of electronic interactions and emergent quantum critical phenomena of itinerant systems. Typically at low enough temperatures and for sufficiently strong interactions, nodal Dirac fermions become susceptible toward a gap opening (mass generation), leading to a maximal gain of the condensation energy. The effective field theory describing the associated quantum phase transition assumes the form of a Gross-Neveu-Yukawa (GNY) model, which besides capturing the dynamics of fermionic and bosonic (order-parameter) fields, also accounts for the Yukawa coupling between them~\cite{zinn-justin}. Traditionally, the GNY theory is analyzed using a perturbative $\epsilon$ expansion, controlled by a parameter $\epsilon=3-d$, measuring the deviation from the upper critical three spatial dimensions. Close to the quantum phase transitions, the notion of sharp fermionic or bosonic excitations becomes moot, and the system accommodates a strongly coupled relativistic (due to an emergent Lorentz symmetry~\cite{anber-lorentz, roy-lorentz, roy-kennett-yang-juricic}) `soup' of these degrees of freedom, constituting a \emph{non-Fermi liquid}.

Such rich field theory predictions recently became relevant in the context of condensed matter physics due to the possible realization of symmetry protected emergent Dirac excitations from concrete lattice models as, for example, in honeycomb~\cite{graphene-review} and $\pi$-flux square~\cite{marston-affleck} lattices. The associated quantum phase transitions are succinctly captured by minimal Hubbardlike models, containing only the finite-range components of the Coulomb interaction~\cite{herbut-solo,raghu-honerkamp,honerpkapm,herbut-juricic-roy, gonzalez, dagofer-hohendler,juricic-roy-TBLG}. Simplicity of these lattice models also permits numerical demonstration of quantum criticality using, for example, quantum Monte Carlo simulations~\cite{sorella-1,herbut-assaad-1,chandrasekharan,troyer-honeycomb,herbut-assaad-2,sorella-2, hong-yao-NN-honeycomb,kaul-itinerant}, besides field theoretic analyses~\cite{rosenstein,herbut-juricic-vafek, herbut-juricic-roy-SC, roy-yang, sslee, hong-yao-1, machiejko-zarf, klebanov, knorr, roy-juricic-PRL}. An intriguing outcome in this context is the following. When Dirac quasiparticles acquire comparable propensity toward the formation of more than one (typically two) competing phases, respectively breaking O$(N_1)$ and O$(N_2)$ symmetries, such that the corresponding order-parameters can be rotated into each other by the generators of the emergent chiral symmetry, then it is conceivable to find a stable itinerant multicritical point (MCP), possessing an enlarged O$(N)$ symmetry, where $N=N_1+N_2$, but $ N \leq 5$ in graphenelike systems, described by an eight-component Dirac spinor~\cite{roy-MCP,royjuricic-MCP,herbut-MCP,roygoswmaijericic-MCP}. In particular, emergent MCPs with O$(N)$ symmetry with $3 \leq N \leq 5$ lack any analogue in pure bosonic systems~\cite{calabrese} and can only be found in strongly interacting Dirac materials.

For the honeycomb lattice model, this observation supports the scenarios with the MCPs of following symmetries: (a) O(3), arising from the competition between Kekul\'{e} valence bond solid (KVBS) or $s$-wave pairing and charge-density-wave, (b) O(4), where KVBS and $s$-wave pairing form a chiral supervector, and (c) O(5), involving anti-ferromagnet and KVBS or quantum spin-Hall insulator and $s$-wave pairing. But, the emergent symmetry at an MCP becomes a subtle issue when one of the competing phases is the discrete three-fold rotational ($Z_3$) symmetry breaking KVBS~\cite{Chamon-Kekule,Chamon_SO5,roy-herbut-Kekule}, since the corresponding cubic term in the bosonic order-parameter theory is \emph{strongly} relevant near $d=3$~\cite{KekuleZ3_1,KekuleZ3_2} and responsible for first-order transitions~\cite{Golner, RMP-Potts}. Therefore, question arises regarding the ultimate stability of the MCPs possessing a seemingly putative enlarged O$(2+N_b)$ symmetry, with $N_b=1,2,3$, respectively, for charge-density-wave, $s$-wave pairing and anti-ferromagnet, in the vicinity of a KVBS ordering. This is the question we address in this work, and arrive at the following conclusions.

\begin{figure}[t!]
\includegraphics[width=.96\linewidth]{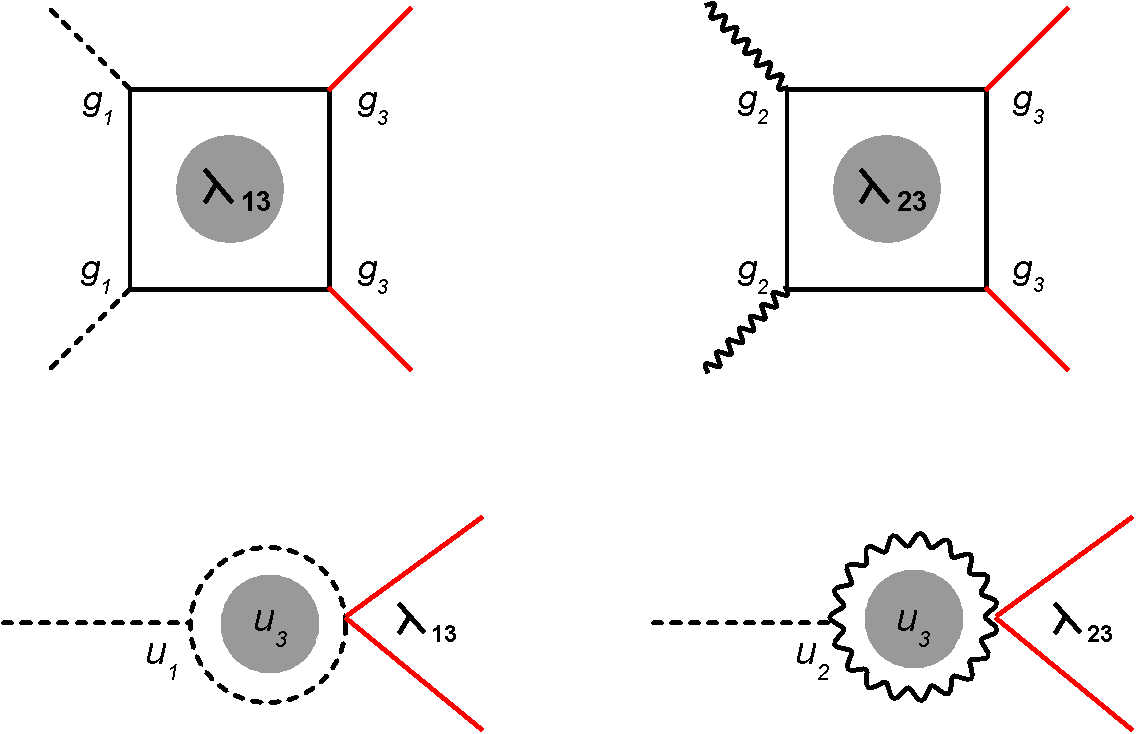}
\caption{Generation of three coupling constants, namely $\lambda_{13}$, $\lambda_{23}$ and $u_{_3}$, appearing in $L_{\rm Mix}$, see Eq.~(\ref{Eq:Lmix}), from the leading order quantum corrections. Yukawa couplings from Eq.~(\ref{Eq:Yukawa}) yield $\lambda_{13}$ and $\lambda_{23}$ (top row), which in turn give rise to the new cubic coupling $u_{_3}$ (bottom row), when combined with $u_{_1}$ and $u_{_2}$ (cubic vertices for Kekul\'e valence bond ordering), see Eq.~(\ref{Eq:kekuleBoson}). Dashed, wavy and red lines respectively represent $\Phi_1$, $\Phi_2$ and ${\boldsymbol \chi}$ bosonic fields, and solid lines to Dirac fermions.
}~\label{Fig:MixedVertices}
\end{figure}

\begin{figure*}[t!]
\includegraphics[width=.32\linewidth]{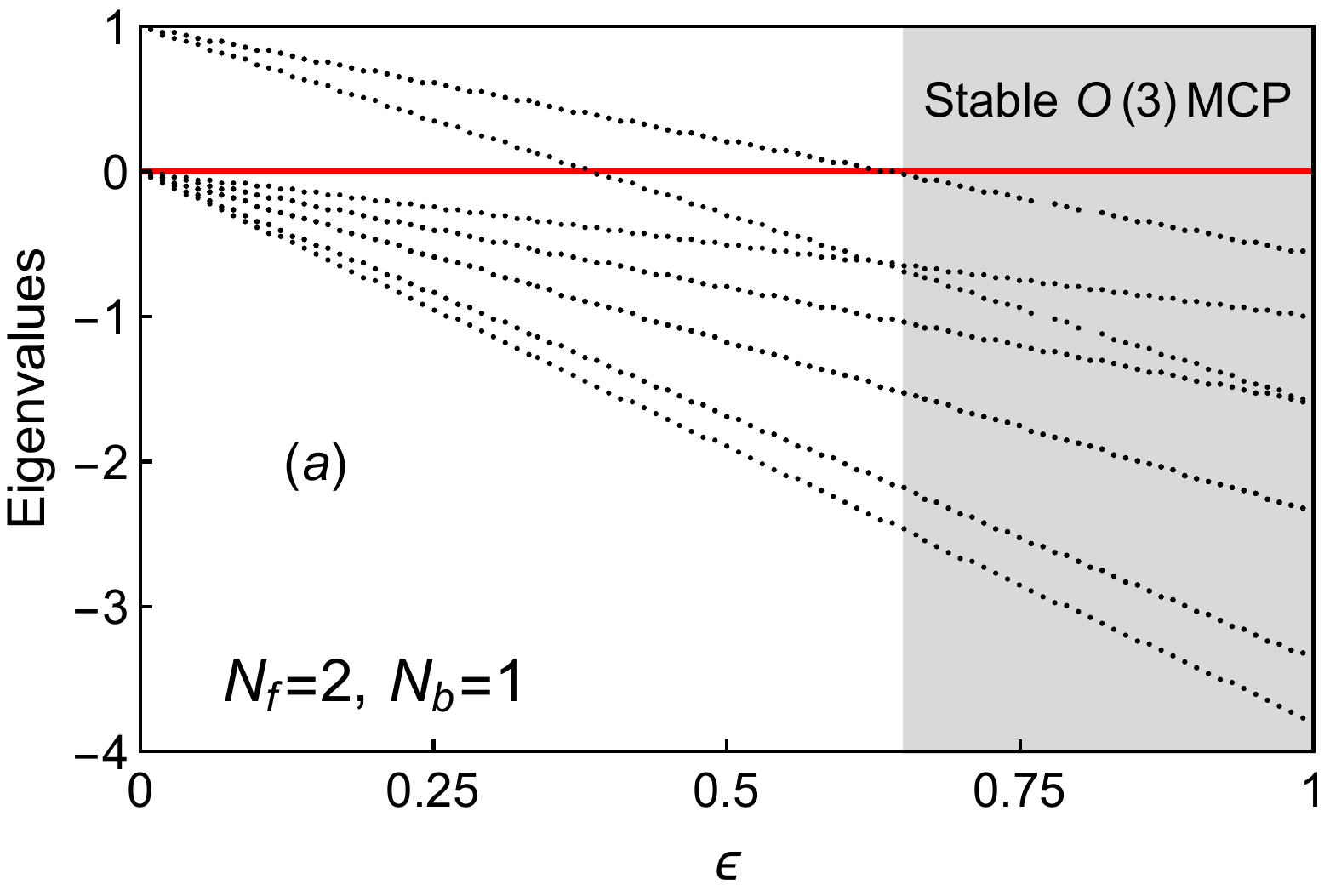}
\includegraphics[width=.32\linewidth]{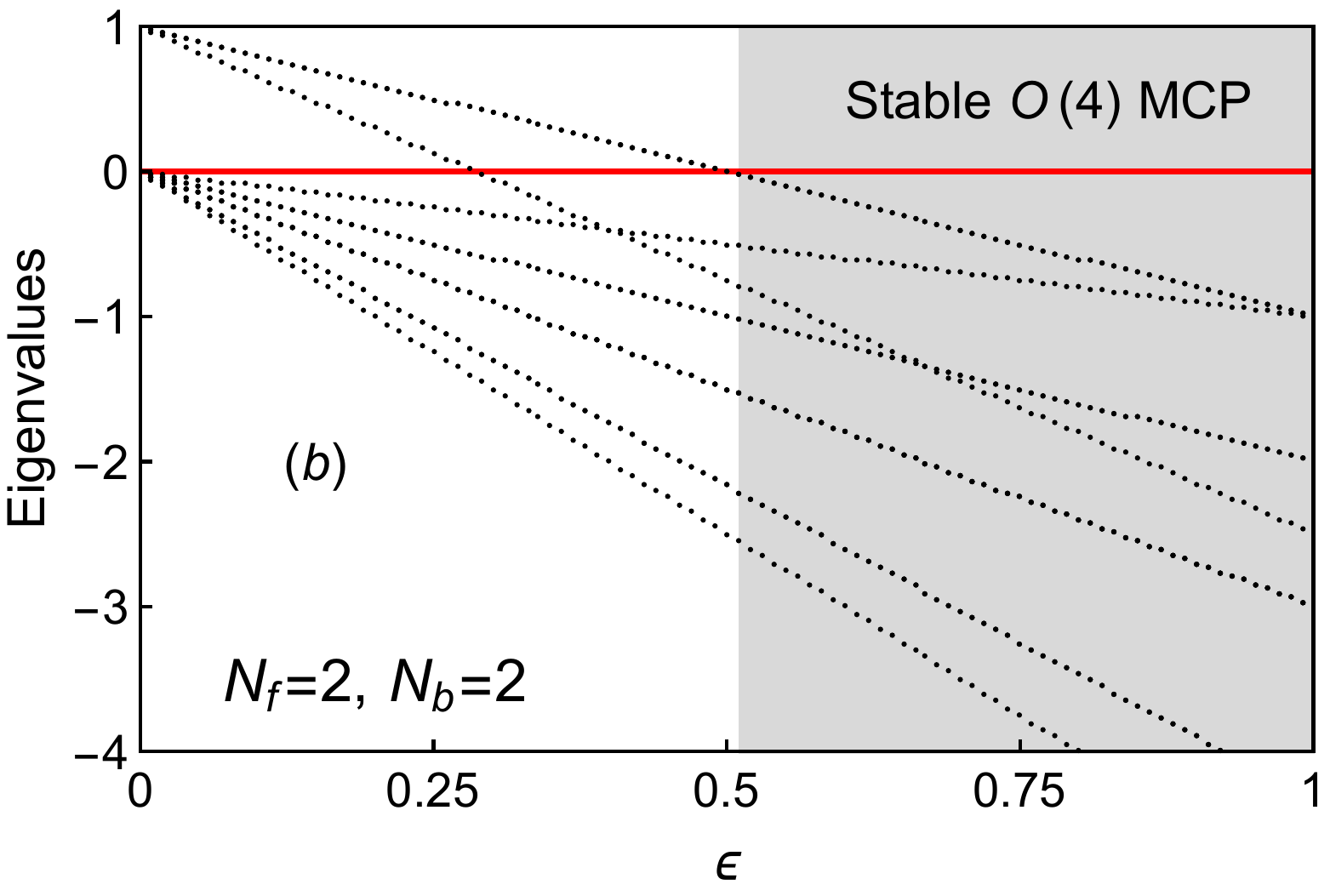}
\includegraphics[width=.32\linewidth]{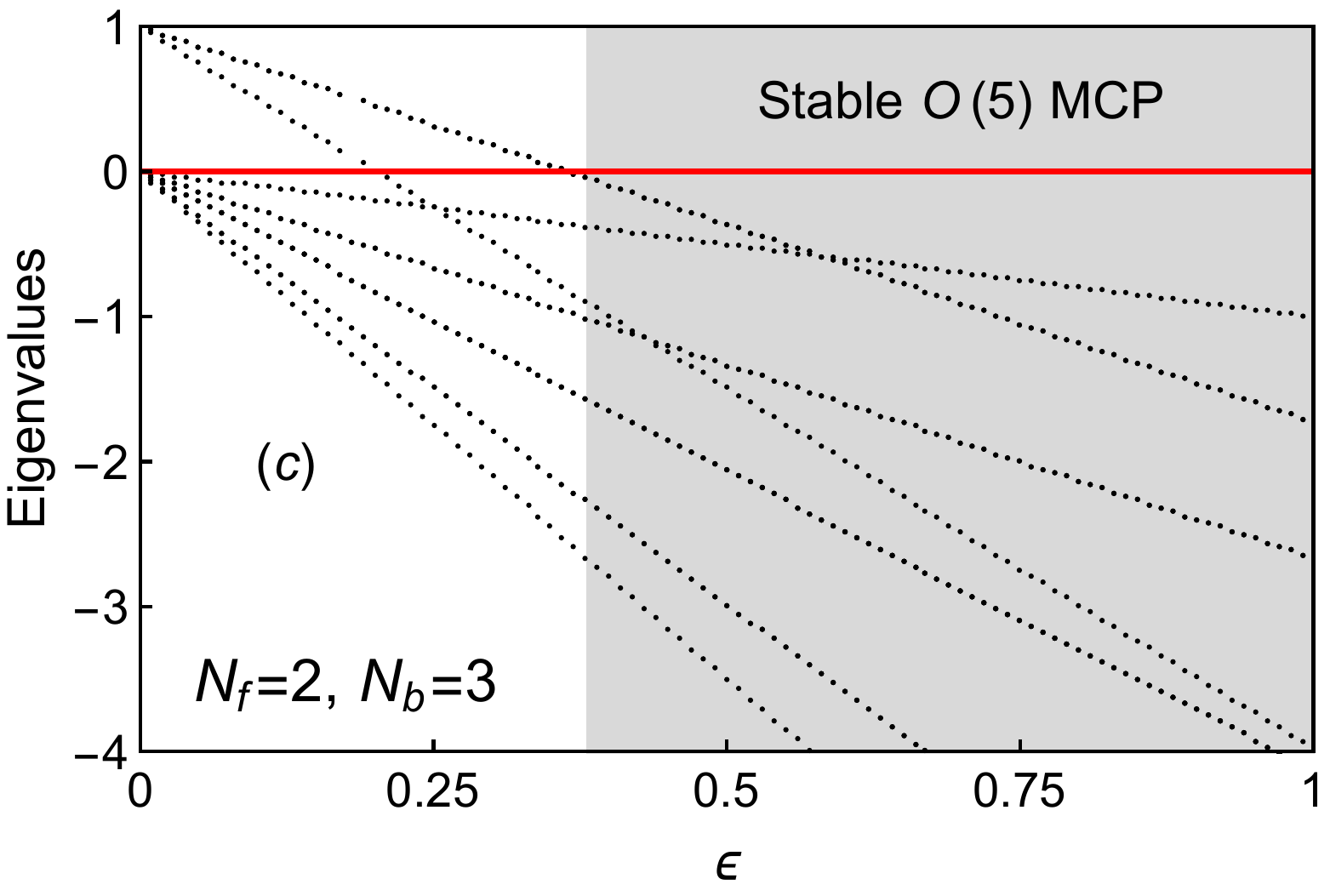}
\caption{ Scaling of the eigenvalues (measured in units of $\epsilon$) of the stability matrix ${\boldsymbol M}$, see Eq.~(\ref{Eq:StabilityMatrix}),  with $\epsilon=3-d$, near O$(2+N_b)$ symmetric multicritical points close to Kekul\'e valence bond solid and (a) charge-density-wave ($N_b=1$), (b) $s$-wave pairing ($N_b=2$) and (c) antiferromagnet ($N_b=3$) orderings. Here, $N_f$ is the number of 4-component spinors and for graphenelike systems $N_f=2$. Close to two spatial dimensions (as $\epsilon \to 1$) all cubic terms become irrelevant (since all eigenvalues of the stability matrix are then negative), implying the ultimate stability of O$(2+N_b)$ symmetric multicritical points. The regime of stability of the multicritical points (the shaded region) increases with the increasing number of the bosonic order parameter components ($N_b$). For any $N_b$, the scaling of 12 eigenvalues collapses onto 7 curves, stemming from the hidden $Z_2 \otimes {\rm O}(N_b+1)$ symmetry of the theory (see text). This feature is also insensitive to the choice of $N_f$ (not shown here explicitly).    
}~\label{Fig:ScalingEVs}
\end{figure*}

We show that when KVBS order, described by a two-component order parameter ${\boldsymbol \Phi}=(\Phi_1,\Phi_2)$, is coupled to an O$(N_b)$ symmetry breaking field (${\boldsymbol \chi}$), a new cubic vertex $\Phi_1 |{\boldsymbol \chi}|^2$ gets generated near a MCP through quantum corrections, besides $\Phi^3_1$ and $\Phi_1 \Phi^2_2$ vertices, see Fig.~\ref{Fig:MixedVertices}. However, due to the Yukawa coupling between gapless fermions and bosons, all the cubic terms ultimately become \emph{irrelevant} close to the O$(2+N_b)$ symmetric MCPs at least near $d=2$, see Fig.~\ref{Fig:ScalingEVs}. Therefore, MCPs with an enlarged O$(2+N_b)$ symmetry near KVBS ordering are stable, and control continuous quantum phase transitions from an interacting Dirac liquid to (1) O$(N_b)$ and (2) O$(N_b+2)$ symmetry breaking insulators, (3) KVBS, as well as (4) the direct transition between KVBS and O$(N_b)$ symmetry breaking insulator.

\emph{Effective field theory:} We begin the discussion by introducing the GNY model in the presence of all symmetry allowed cubic terms close to an $O(2+N_b)$ symmetric quantum MCP. The corresponding imaginary time ($\tau$) action reads as ${\mathcal S}= \int d\tau d^d {\boldsymbol r} L$, with $L=L_f+L_{bf}+L_b$ and ${\boldsymbol r}= \left( x_1, \cdots, x_d \right)$ is the spatial coordinate. The dynamics of massless Dirac fermions is captured by 
\begin{align}
L_f= \Psi^\dagger (\tau, {\bf r}) \bigg[ \partial_\tau - i \sum^{d}_{j=1} \Gamma_j \partial_j \bigg] \Psi (\tau, {\bf r}),
\end{align}
where $\Gamma_j$s are mutually anticommuting eight-dimensional Hermitian matrices, and $\Psi^\dagger$ and $\Psi$ are independent eight-component Grassmann variables. The following discussion is, however, impervious to specific matrix and spinor representations. The coupling between massless Dirac fermions and bosonic order parameter fields reads 
\begin{align}~\label{Eq:Yukawa}
L_{bf} = \sum^2_{j=1} g_{_j} \Phi_j \Psi^\dagger M_j \Psi 
+ g_{_3} \sum^{N_b}_{k=1} \chi_{_k} \Psi^\dagger M_{2+k} \Psi, 
\end{align} 
where $X \equiv X(\tau,{\bf r})$ for $X=\Psi^\dagger, \Psi, {\boldsymbol \Phi}, {\boldsymbol \chi}$. Mutually anticommuting eight-dimensional Hermitian mass matrices $M_j$s, satisfy $\{ \Gamma_i , M_j \}=0$, with $N_b \leq 5-d$. The $O(2+N_b)$ chiral rotations among the mass matrices are generated by $G_{ij}=[M_i,M_j]/(2i)$, where $i,j=1, \cdots, N_b +2$. Note $G_{ij}$s are the generators of the chiral symmetry for massless Dirac fermions as $[\Gamma_i, G_{jk}]=0$.

The pure bosonic part of theory can be decomposed according to $L_b=L_{\rm Kek} + L^{N_b}_{\rm Sym}+ L_{\rm Mix}$, with 
\allowdisplaybreaks[4]
\begin{align}
L_{\rm Kek} &= \sum^2_{j=1} \left[ \frac{1}{2} \left( \partial_\mu \Phi_j \right)^2 +m^2_j \Phi^2_j \right] + \frac{u_{_1}}{3!} \Phi^3_1 + \frac{u_{_2}}{2!} \Phi_1 \Phi^2_2 \nonumber \\
& + \sum^2_{j=1}\frac{\lambda_j}{4!} \Phi^4_j + \frac{2 \lambda_{12}}{4!} \Phi^2_1 \Phi^2_2,~\label{Eq:kekuleBoson} \\
L^{N_b}_{\rm Sym} &= \sum^{N_b}_{j=1} \left[ \frac{1}{2} \left( \partial_\mu \chi_j \right)^2 +m^2_3 \chi^2_j + \frac{\lambda_3}{4!} \left( \chi^2_j \right)^2 \right], \\
L_{\rm Mix} &= \frac{u_{_3}}{2!} \Phi_1 \sum^{N_b}_{j=1} \chi^2_j + \left( \sum^2_{j=1}\frac{2 \lambda_{j3} }{4!} \; \Phi^2_j \right) \; \sum^{N_b}_{k=1} \chi^2_j.~\label{Eq:Lmix}
\end{align}
Therefore, the effective field theory contains 12 coupling constants. Even though three coupling constants appearing in $L_{\rm Mix}$ are absent at the bare level, $u_{_3}$, $\lambda_{13}$ and $\lambda_{23}$ get generated through quantum corrections, see Fig.~\ref{Fig:MixedVertices}, and thus have to be included from the outset.

Near a pure KVBS ordering $L^{N_b}_{\rm Sym}= L_{\rm Mix}=0$ and $g_{_3}=0$. In addition, $u_1=-u_2$, $g_{_1}=g_{_2}$ and $\lambda_1=\lambda_2=\lambda_{12}$. However, such a symmetry is broken, when the system acquires comparable propensity toward the formation of an O$(N_b)$ symmetry breaking phase. Nevertheless, the above effective field theory enjoys a hidden O$(N_b+1)$ symmetry, which we identify by constructing a composite bosonic field according to $(\Phi_2, {\boldsymbol \chi})$, and setting $g_{_2}=g_{_3}$, $u_{_2}=u_{_3}$, $\lambda_2=\lambda_3=\lambda_{23}$ and $\lambda_{12}=\lambda_{13}$. The effective field theory then describes a $Z_2 \otimes {\rm O}(N_b+1)$ symmetric GNY model for massless Dirac fermions in the presence of \emph{cubic terms} and contains 7 coupling constants. The renormalization group (RG) flow equations also reflect this symmetry, about which more in a moment. Anticipating the outcome, we set the Fermi ($v_{_F}$) and bosonic ($v_{_B}$) velocities to be equal (due to an emergent Lorentz symmetry at MCP)~\cite{roy-lorentz}, and $v_{_F}=v_{_B}=1$ for simplicity.

\emph{$\epsilon$ expansion}: The bare (engineering) scaling dimension of the fermionic and bosonic fields, respectively,  $D[\Psi]=d/2$ and $D[{\boldsymbol \Phi}]=D[{\boldsymbol \chi}]=(d-1)/2$, follow from the corresponding noninteracting parts of the action, the scaling dimension of momentum $D[k]=1$ and imaginary time $D[\tau]=-1$. Consequently, the scaling dimension of all three Yukawa couplings and six four-boson couplings is $D[\{g^2\}]=D[\{\lambda\}]=3-d$, and that for the bosonic mass parameters (controlling the transition from a Dirac semimetal to ordered phases) $D[\{ m^2 \}]=2$~\cite{footnote}. Our focus here is restricted on the critical hyperplane, defined by $\{ m^2 \}=0$. Therefore, in the absence of any cubic terms, the GNY model near fermionic MCPs can be addressed in terms of a perturbative $\epsilon$ expansion about three spatial dimensions, with $\epsilon=3-d$~\cite{roy-MCP,royjuricic-MCP,herbut-MCP,roygoswmaijericic-MCP}.

On the other hand, the scaling dimension of all three cubic terms is $D[\{ u \}]=(2+\epsilon)/2$, implying that they are \emph{strongly} relevant perturbations at the bare level in $d=3$~\cite{KekuleZ3_1,KekuleZ3_2}. Therefore, the entire theory cannot be controlled in terms of a single expansion parameter $\epsilon$. Alternatively, one may pursue a \emph{double} $\epsilon$ expansion, with $D[\{g^2\}]=D[\{\lambda\}]=\epsilon_1$ and $D[\{ u \}]=\epsilon_2/2$, where $\epsilon_1=3-d$ and $\epsilon_2=5-d$, somewhat similar to the ones employed in interacting and disordered systems~\cite{boyanovsky-cardy,roy-dassarma,yerzhakov-maciejko}. However, such double $\epsilon$ expansion does not yield any additional control over the perturbative analysis. So, we abandon it for rest of the discussion. Nonetheless, the cubic vertices receive perturbative corrections, and the question we seek to answer in rest of the paper is whether these corrections (in particular, the ones arising due to the Yukawa couplings) can provide sufficiently \emph{negative} scaling dimensions to the cubic vertices to turn them \emph{irrelevant} near the fermionic MCPs, located at $\{ g^2\} \sim \epsilon$, $\{ \lambda \} \sim \epsilon$ and $\{ u \}=0$, at least when $\epsilon \to 1$ or as $d \to 2$.

\emph{RG analysis}: To answer this question, we here restrict ourselves to the leading order $\epsilon$ expansion for $\{ g \}$ and $\{ \lambda \}$, and account for one-loop corrections to $\{ u \}$. To this end we integrate over the Matsubara frequency $-\infty \leq \omega \leq \infty$ and the fast Fourier modes, living within a thin Wilsonian momentum shell $\Lambda e^{-\ell} \leq |{\bf k}| \leq \Lambda$. We perform the matrix algebra in $d=2$ and subsequently carry the momentum integral in dimension $d=3-\epsilon$. The resulting coupled RG flow equations are 
\begin{widetext}
\allowdisplaybreaks[4]
\begin{align} 
\frac{dg^2_{_1}}{d \ell} &=\epsilon g^2_{_1} -\left( 2 N_f +3 \right) g^4_{_1} + g^2_{_1} \left(g^2_{_2}+ N_b g^2_{_3} \right) 
-\frac{1}{6} \left( u^2_{_1} + u^2_{_2} + N_b u^2_{_3} \right) g^2_{_1}, \nonumber \\
\frac{dg^2_{_2}}{d \ell} &=\epsilon g^2_{_2} -\left( 2 N_f +3 \right) g^4_{_2} + g^2_{_2} \left(g^2_{_1}+ N_b g^2_{_3} \right)
-\frac{1}{3} u^2_{_2} g^2_{_2}, 
\quad 
\frac{dg^2_{_3}}{d \ell} =\epsilon g^2_{_3} - \left( 2 N_f +4 -N_b \right) g^4_{_3} + g^2_{_3} \left( g^2_{_1} + g^2_{_2}\right)
-\frac{1}{3} u^2_{_3} g^2_{_3}, \nonumber \\
\frac{du_{_1}}{d \ell} &=\frac{2+\epsilon}{2}u_{_1}-3 N_f g^2_{_1} u_{_1} + 
\frac{1}{4} \left[ 11 u^3_{_1} + 12 u^3_{_2} + 12 u^3_{_3} - u_{_1} \left( u^2_{_2} + N_b u^2_{_3} \right) \right]
-\frac{1}{2} \left[ 3 u_{_1} \lambda_1 +u_{_2} \lambda_{12} + N_b u_{_3} \lambda_{13}  \right], \nonumber \\
\frac{du_{_2}}{d \ell} &=\frac{2+\epsilon}{2}u_{_2}-N_f \left( g^2_{_1} + 2 g^2_{_2}\right) u_{_2}
+ \frac{u_{_2}}{12} \left[ 7 u^2_{_2} + 12 u_{_1} u_{_2} - u^2_1 - N_b u^2_{_3} \right]
-\frac{1}{3} \left[3 u_2 \lambda_2 + \left(u_{_1}+u_{_2} \right) \lambda_{12} + N_b u_{_3} \lambda_{23} \right], \nonumber  \\
\frac{du_{_3}}{d \ell} &=\frac{2+\epsilon}{2} u_{_3} - N_f \left( g^2_{_1} + 2 g^2_{_3} \right) u_{_3} 
+ \frac{u_{_3}}{12} \left[ \left( 8-N_b \right) u^2_{_3} +12 u_{_1} u_{_3} - u^2_{_1} - u^2_{_2} \right]
-\frac{1}{3} \left[ \left( N_b+2 \right) u_{_3} \lambda_3 + \left(u_{_1}+ u_{_3} \right) \lambda_{13} + u_{_2} \lambda_{23}  \right], \nonumber \\
\frac{d \lambda_3}{d \ell} &=\epsilon \lambda_3 - 4 N_f g^2_{_3} \left( \lambda_3 -6 g^2_{_3} \right)-\frac{1}{6} \left[ \left( N_b+8 \right)\lambda^2_3 + \lambda^2_{13} + \lambda^2_{23} \right] + u^2_{_3} \left[\frac{34}{3} \lambda_3 + 4 \lambda_{13} \right] - 24 u^4_{_3}, \nonumber \\
\frac{d\lambda_1}{d \ell} &= \epsilon \lambda_1 - 4 N_f g^2_{_1} \left( \lambda_1 -6 g^2_{_1} \right)-\frac{1}{6} \left( 9 \lambda^2_1 +\lambda^2_{12}+N_b \lambda^2_{13} \right) + 4 \left( 3 u^2_{_1} \lambda_1 + u^2_{_2} \lambda_{12} + N_b u^2_{_3} \lambda_{13}  \right) 
-\frac{1}{3} \left( u^2_{_1} +u^2_{_2} + N_b u^2_{_3} \right) \lambda_1 \nonumber \\
&- 12 \left( u^4_{_1} + u^4_{_2} + N_b u^4_{_3} \right), \nonumber \\
\frac{d \lambda_2}{d \ell} &= \epsilon \lambda_2 - 4 N_f g^2_{_2} \left( \lambda_2 - 6 g^2_{_2}\right) 
-\frac{1}{6} \left( 9 \lambda^2_2 +\lambda^2_{12} + N_b \lambda^2_{23} \right)
+u^2_{_2} \left[ \frac{34}{3} \lambda_2 + 4 \lambda_{12} \right] -24 u^4_{_2}, \nonumber \\
\frac{d \lambda_{12}}{d \ell} &= \epsilon \lambda_{12} -2 N_f \left( g^2_{_1} + g^2_{_2} \right) \lambda_{12} + 24 N_f g^2_{_1} g^2_{_2}
-\frac{1}{6} \left[ 3 \lambda_{12} \left( \lambda_1+\lambda_2 \right) + 4 \lambda^2_{12} + N_b \lambda_{13} \lambda_{23} \right] \nonumber \\
&+ \frac{1}{6} \left( 11 u^2_{_1} +33 u^2_{_2} +24 u_{_1} u_{_2} -N_b u^2_{_3} \right) \lambda_{12} + 6 u^2_{_2} \left( \lambda_1 + \lambda_2 \right) + 2 N_b u^2_{_3} \lambda_{23} -24 \left( u^4_{_2} + u^2_{_1} u^2_{_2}  + u_{_1} u^3_{_2} \right), \nonumber \\
\frac{d \lambda_{13}}{d \ell} &= \epsilon \lambda_{13} - 2 N_f \left( g^2_{_1} + g^2_{_3} \right) \lambda_{13} + 24 N_f g^2_{_1} g^2_{_3} -\frac{1}{6} \left[ \lambda_{13} \left( 3 \lambda_1 + \left( N_b +2 \right) \lambda_3 + 4 \lambda_{13} \right) + \lambda_{12} \lambda_{23} \right] 
-\frac{\lambda_{13}}{6} \left[ u^2_{_1} + u^2_{_2} + \left( N_b +2 \right) u^2_{_3} \right] \nonumber \\
&+ 2  \left[ \left( N_b+2 \right) u^2_{_3} \lambda_3 + u^2_{_1} \lambda_{13} +  u^2_{_2} \lambda_{23} +  3 u^2_{_3} \left( \lambda_1 + \lambda_{13} \right)+ 2 u_{_1} u_{_3} \lambda_{13}  \right] - 24 \left( u^2_{_1} u^2_{_3} +u_{_1} u^3_{_3} + u^4_{_3} \right), \nonumber \\
\frac{d \lambda_{23}}{d \ell} &= \epsilon \lambda_{23} - 2 N_f \left( g^2_{_2} + g^2_{_3} \right) \lambda_{23} + 24 N_f g^2_{_2} g^2_{_3} -\frac{1}{6} \left[ 3 \lambda_2 \lambda_{23} + \left( N_b + 2 \right) \lambda_3 \lambda_{23} + 4 \lambda^2_{23} + \lambda_{12} \lambda_{13} \right] -\frac{1}{3} \left( u^2_{_2} + u^2_{_3} \right) \lambda_{23} \nonumber \\
&+  \left[ 2 \left(u^2_{_2} \lambda_{13} + u^2_{_3} \lambda_{12} \right) + 3 \left( u_{_2} + u_{_3} \right)^2 \lambda_{23} \right] -24 u^2_{_2} u^2_{_3},
\end{align}
\end{widetext}
in terms of dimensionless coupling constants, defined as $X \Lambda^{-\epsilon}/(8\pi^2) \to X$, where $X= \{ g^2 \}, \{\lambda\}$ and $\{ u \} \Lambda^{-\frac{2+\epsilon}{2}}/(8 \pi^2) \to \{ u \}$. Here, $N_f$ is the number of four-component fermion flavors, and hence for graphenelike systems $N_f=2$. The details of the RG calculation are presented in the Supplemental Materials~\cite{supple}. The underlying $Z_2 \otimes {\rm O}(N_b+1)$ symmetry of the GNY theory can be appreciated in the following way. If we set $g_{_2}=g_{_3}$, $u_{_2}=u_{_3}$, $\lambda_2=\lambda_3=\lambda_{23}$ and $\lambda_{12}=\lambda_{13}$, then RG flow equations of the following couplings are identical:  (1) $g_{_2}$ and $g_{_3}$, (2) $u_{_2}$ and $u_{_3}$, (3) $\lambda_2$, $\lambda_3$ and $\lambda_{23}$, and (4) $\lambda_{12}$ and $\lambda_{13}$. It is quite challenging to find all possible solutions of above 12 coupled RG flow equations, so we instead focus on the specific and relevant case, the O$(2+N_b)$ symmetric MCP, located at $g^2_{_1}=g^2_{_2}=g^2_{_3}=g^2_\ast$, $\lambda_1=\lambda_2=\lambda_{12}=\lambda_3=\lambda_{13}=\lambda_{23}=\lambda_\ast$, and $u_{_1}=u_{_2}=u_{_3}=u_\ast$, where 
\begin{align}
\left( g^2_\ast, \lambda_\ast, u_\ast \right) = \left( \epsilon, 3 \; \frac{H(N_f,N_b)}{10+N_b} \epsilon,0 \right) \frac{1}{2 (N_f +1) -N_b},
\end{align}    
and $H(x,y)=2-2x-y + [4 x^2 + (y-2)^2+ 4 x (38 +5 x)]^{1/2}$.

To analyze the stability of such fixed points, we compute the stability matrix (${\boldsymbol M}$), defined as 
\begin{align}~\label{Eq:StabilityMatrix}
M_{ij} \left( \{ C\} \right)= \frac{d}{d C_j} \left( \frac{d C_i}{d \ell}\right),
\end{align}
 and its eigenvalues in its vicinity. Here $\{ C \}$ is the set of 12 coupling constants, and thus $i,j=1, \cdots, 12$. The results are displayed in Fig.~\ref{Fig:ScalingEVs}. Sufficiently close to two spatial dimensions (as $\epsilon \to 1$), all 12 eigenvalues of the stability matrix are negative for any value of $N_f$ and $N_b$. Hence, all cubic terms (namely, $u_{_1},u_{_2}$ and $u_{_3}$) become irrelevant in the close vicinity of the O$(2+N_b)$ symmetric MCPs, indicating their stability in two dimensions. Note that irrelevance of the cubic terms is solely introduced by nontrivial Yukawa coupling between gapless bosonic and fermionic degrees of freedom. Hence, such a quantum MCP can only be realized in strongly interacting Dirac systems. We also note that with increasing number of the order-parameter components $N_b$, the regime of stability (the shaded region in Fig.~\ref{Fig:ScalingEVs}) of the MCPs increases~\cite{footnote_2d}. The fact that the MCPs are stable over a range of $\epsilon$ (see Fig.~\ref{Fig:ScalingEVs}), suggests that their stability in the presence of cubic couplings is possibly nonperturbative in nature. Only the range of $\epsilon$, over which the MCPs are stable, can be renormalized at each order in a perturbative expansion.

We should also mention that a leading order $\epsilon$ expansion in a purely bosonic system suggests a putative O(3) symmetric MCP, which, however, looses stability once the higher order corrections are taken into account~\cite{calabrese}. On the other hand, there exists neither O(4) nor O(5) symmetric MCP in a purely bosonic theory. Therefore, appearance of O$(2+N_b)$ symmetric quantum MCPs, with $N_b=1,2,3$, and their stability against the cubic perturbations, when the system resides at the brink of Kekul\'e ordering in honeycomb lattice, are purely fermion driven phenomena in Dirac materials. Also a pure Kekul\'e O$(2)$ quantum critical point (in the absence of the ${\boldsymbol \chi}$ field) is stabilized due to gapless Dirac fermions~\cite{KekuleZ3_1, KekuleZ3_2, comment-Kekule, footnote_rosenstein, footnote_2d}.

Besides the O$(2+N_b)$ symmetric MCP, there exists two more interacting fixed points, possessing $O(2)$ and $O(N_b)$ symmetries. Respectively, they control transition from a Dirac semimetal to KVBS and O$(N_b)$ symmetry breaking phase. The cubic terms are irrelevant at these two fixed points. However, they ultimately become unstable toward the MCP. We could not find (numerically) any fixed point at finite $\{ u_i \}$, when $\epsilon$ is close to 1.

\emph{Discussion}: To summarize, we address the stability of quantum MCPs with enlarged $O(2+N_b)$ symmetry, when a correlated Dirac liquid, realized on a honeycomb lattice, acquires comparable propensity toward the nucleation of KVBS and charge-density-wave ($N_b=1$) or $s$-wave pairing ($N_b=2$) or antiferromagnet ($N_b=3$). We show that quantum corrections generate new cubic vertices (see Fig.~\ref{Fig:MixedVertices}) near such MCPs, besides the ones for KVBS due to the breaking of discrete $Z_3$ rotational symmetry in honeycomb lattice. All cubic terms are strongly relevant at the bare level and responsible for a generic first-order transition in pure bosonic systems~\cite{Golner,RMP-Potts}. However, due to nontrivial boson-fermion Yukawa couplings, they all become irrelevant near high symmetric MCPs, close to two spatial dimensions, see Fig.~\ref{Fig:ScalingEVs}. Therefore, $O(2+N_b)$ symmetric quantum MCPs are expected to be stable in honeycomb Dirac systems. At this MCP the bosonic and fermionic anomalous dimensions are respectively given by $\eta_b= 2 N_f g^2_\ast$ and $\eta_\Psi = (2+N_b) g^2_\ast/2$, and the correlation length exponent is 
\begin{equation}
\nu=\frac{1}{2} + \frac{N_f}{2} g^2_\ast + \frac{4+N_b}{24} \lambda_\ast. \nonumber 
\end{equation} 
Together they determine the universality class of continuous transitions from a Dirac semimetal to (a) KVBS, (b) $O(N_b)$ and (c) $O(2+N_b)$ symmetry breaking orders, and (d) the direct transition from KVBS to an $O(N_b)$ symmetry breaking order through itinerant MCP.

Our results should be germane in the context of recent quantum Monte Carlo simulations on correlated Dirac liquid, in the presence of competing orderings~\cite{sato-hohendler-assaad, HongYao-2019}. So far, emergence of high symmetry has only been reported slightly away from the itinerant MCP. We hope that our results will motivate future works to explore symmetry enlargement in the proximity to the MCP and KVBS ordering. In addition, our results can also be relevant for slow (due to sufficiently small Fermi velocity) and strongly interacting (due to substantial bandwidth suppression) Dirac fermions in twisted bilayer graphene near magic angle (MA-TBLG)~\cite{TBLG-1, TBLG-2}, and in correlated organic Dirac material pressured $\alpha$-(BEDT-TTF)$_2$I$_3$~\cite{organicDirac}. A recent experiment reported the existence of an insulating phase near the charge-neutrality point in the former system~\cite{TBLG-3}. With the application of suitable non-thermal tuning parameters (for example, pressure, strain, twist angle etc.) it is, at least in principle, conceivable to drive such a Dirac insulator through a quantum MCP, where the present discussion can be pertinent~\cite{Kekule-MA-TBLG}.


\begin{thebibliography}{}

\bibitem{zinn-justin} J. Zinn-Justin, \emph{Quantum Field Theory and Critical Phenomena} (Oxford University Press, Oxford, UK, 2002).


\bibitem{anber-lorentz} M. M. Anber and J. F. Donoghue, Phys. Rev. D {\bf 83}, 105027 (2011).

\bibitem{roy-lorentz} B. Roy, V. Juri\v ci\' c and I. F. Herbut, JHEP {\bf 04}, (2016) 018.

\bibitem{roy-kennett-yang-juricic} B. Roy, M. P. Kennett, K. Yang and V. Juri\v ci\' c, Phys. Rev. Lett. {\bf 121}, 157602 (2018). 


\bibitem{graphene-review}  A. H. Castro Neto, F. Guinea, N. M. R. Peres, K. S. Novoselov, A. K. Geim, Rev. Mod. Phys. {\bf 81}, 109 (2009).

\bibitem{marston-affleck} I. Affleck and J. B. Marston, Phys. Rev. B {\bf 37}, 3774 (1988).


\bibitem{herbut-solo} I. F. Herbut, Phys. Rev. Lett. {\bf 97}, 146401 (2006).

\bibitem{raghu-honerkamp} S. Raghu, X-L. Qi, C. Honerkamp, S-C. Zhang, Phys. Rev. Lett. {\bf 100}, 156401 (2008).

\bibitem{honerpkapm} C. Honerkamp, Phys. Rev. Lett. {\bf 100}, 146404 (2008).

\bibitem{herbut-juricic-roy} I. F. Herbut, V. Juri\v ci\' c, and B. Roy, Phys. Rev. B {\bf 79}, 085116 (2009).

\bibitem{gonzalez} J. Gonzalez, JHEP {\bf 07}, 175 (2013).

\bibitem{dagofer-hohendler} M. Daghofer and M. Hohenadler, Phys. Rev. B {\bf 89}, 035103 (2014).

\bibitem{juricic-roy-TBLG} B. Roy and V. Juri\v ci\' c, Phys. Rev. B {\bf 99}, 121407 (2019).


\bibitem{sorella-1} S. Sorella, Y. Otsuka, and S. Yunoki, Sci. Rep. {\bf 2}, 992 (2012).

\bibitem{herbut-assaad-1} F. F. Assaad and I. F. Herbut, Phys. Rev. X {\bf 3}, 031010 (2013).

\bibitem{chandrasekharan} S. Chandrasekharan and A. Li, Phys. Rev. D {\bf 88}, 021701(R) (2013).

\bibitem{troyer-honeycomb} L. Wang, P. Corboz, and M. Troyer, New J. Phys. {\bf 16}, 103008 (2014).

\bibitem{herbut-assaad-2} F. P. Toldin, M. Hohenadler, F. F. Assaad, and I. F. Herbut, Phys. Rev. B {\bf 91}, 165108 (2015).

\bibitem{sorella-2} Y. Otsuka, S. Yunoki, and S. Sorella, Phys. Rev. X {\bf 6}, 011029 (2016).

\bibitem{hong-yao-NN-honeycomb} Z-X. Li, Y-F. Jiang, and H. Yao, New J. Phys. {\bf 17}, 085003 (2015).

\bibitem{kaul-itinerant} S. Pujari, T. C. Lang, G. Murthy, R. K. Kaul, Phys. Rev. Lett. {\bf 117}, 086404 (2016).


\bibitem{rosenstein} B. Rosenstein and A. Kovner, Phys. Lett. B {\bf 314}, 381 (1993).

\bibitem{herbut-juricic-vafek} I. F. Herbut, V. Juri\v ci\' c, and O. Vafek, Phys. Rev. B {\bf 80}, 075432 (2009).

\bibitem{herbut-juricic-roy-SC} B. Roy, V. Juri\v ci\' c and I. F. Herbut, Phys. Rev. B {\bf 87}, 041401(R) (2013); Erratum: Phys. Rev. B {\bf 94}, 119901 (2016).

\bibitem{roy-yang} B. Roy, and K. Yang, Phys. Rev. B {\bf 88}, 241107(R) (2013).

\bibitem{sslee} P. Ponte and S-S. Lee, New J. Phys. {\bf 16}, 013044 (2014).

\bibitem{hong-yao-1} S-K. Jian, Y-F. Jiang, and H. Yao, Phys. Rev. Lett. {\bf 114}, 237001 (2015).

\bibitem{machiejko-zarf} N. Zerf, C-H Lin, and J. Maciejko, Phys. Rev. B {\bf 94}, 205106 (2016).

\bibitem{klebanov} L. Fei, S. Giombi, I. F. Klebanov, G. Tarnopolsky, Prog. Theor. Exp. Phys. {\bf 2016}, 12C105.

\bibitem{knorr} B. Knorr, Phys. Rev. B {\bf 94}, 245102 (2016).

\bibitem{roy-juricic-PRL} B. Roy and V. Juri\v ci\' c, Phys. Rev. Lett. {\bf 121}, 137601 (2018).


\bibitem{roy-MCP} B. Roy,  Phys. Rev. B {\bf 84}, 113404 (2011).  

\bibitem{royjuricic-MCP} B. Roy and V. Juri\v ci\' c, Phys. Rev. B {\bf 90}, 041413 (2014). 

\bibitem{herbut-MCP} L. Janssen, I. F. Herbut, M. M. Scherer, Phys. Rev. B {\bf 97}, 041117 (2018).  

\bibitem{roygoswmaijericic-MCP} B. Roy. P. Goswami and V. Juri\v ci\' c, Phys. Rev. B {\bf 97}, 205117 (2018).

\bibitem{calabrese} P. Calabrese, A. Pelissetto, and E. Vicari, Phys. Rev. B {\bf 67}, 054505 (2003).
 

\bibitem{Chamon-Kekule} C.-Y. Hou, C. Chamon, and C. Mudry, Phys. Rev. Lett. {\bf 98}, 186809 (2007).

\bibitem{Chamon_SO5} S. Ryu, C. Mudry, C.-Y. Hou, and C. Chamon, Phys. Rev. B {\bf 80}, 205319 (2009).

\bibitem{roy-herbut-Kekule}  B. Roy and I. F. Herbut, Phys. Rev. B {\bf 82}, 035429 (2010).


\bibitem{KekuleZ3_1} Z.-X. Li, Y.-F. Jiang, S.-K. Jian, and H. Yao, Nat. Commun. {\bf 8}, 314 (2017). 

\bibitem{KekuleZ3_2} E. Torres, L. Classen, I. F. Herbut, M. M. Scherer,  Phys. Rev. B {\bf 97}, 125137 (2018).


\bibitem{Golner} G. R. Golner, Phys. Rev. B {\bf 8}, 3419 (1973).

\bibitem{RMP-Potts} F. Y. Wu, Rev. Mod. Phys. {\bf 54}, 235 (1982).


\bibitem{footnote} The sets of 3 Yukawa, 6 four-boson and 3 cubic couplings are respectively denoted by $\{ g^2\}$, $\{ \lambda \}$ and $\{ u \}$. The set of three bosonic masses is represented by $\{ m^2 \}$.


\bibitem{boyanovsky-cardy} D. Boyanovsky and J. L. Cardy, Phys. Rev. B {\bf 26}, 154(1982).

\bibitem{roy-dassarma} B. Roy, and S. Das Sarma, Phys. Rev. B {\bf 94}, 115137 (2016).

\bibitem{yerzhakov-maciejko} H. Yerzhakov, and J. Maciejko, Phys. Rev. B {\bf 98}, 195142 (2018).


\bibitem{supple} See Supplemental Materials at XXX-XXXX for detailed derivation of the leading order RG flow equations.


\bibitem{footnote_2d} Four-fermion theory for spinless fermions also supports an O$(3)$ symmetric critical point, where the KVBS and charge-density-wave orders can be chirally rotated into each other, as found from an $\epsilon$ expansion about the lower critical one dimension, with $\epsilon=d-1$~\cite{herbut-juricic-roy}. However, it does not account for the order-parameter fluctuations and the effects of $Z_3$ symmetry breaking in the KVBS. Possible demonstration of an O$(5)$ symmetric critical point for spinful fermions will be addressed in future.


\bibitem{comment-Kekule} The O(2) Kekul\'{e} critical point is stable only for $\epsilon>0.75$ when $N_f=2$ (obtained from leading order RG calculation). Therefore, O$(2+N_b)$ symmetric itinerant MCPs are more stable than the pure O(2) Kekul\'{e} critical point.


\bibitem{footnote_rosenstein} An Ising GNY critical point is stable in the presence of cubic interaction terms, as found from a large-$N_f$ analysis (but neglecting the boson dynamics) at $d=2$. See G. Gat, A. Kovner and B. Rosenstein, Nucl. Phys. B {\bf 385}, 76 (1992).





\bibitem{sato-hohendler-assaad} T. Sato, M. Hohenadler, F. F. Assaad, Phys. Rev. Lett. {\bf 119}, 197203 (2017).

\bibitem{HongYao-2019} Z-X. Li, S-K. Jian, and H. Yao, arXiv:1904.10975

\bibitem{TBLG-1} Y. Cao, V. Fatemi, A. Demir, S. Fang, S. L. Tomarken, J. Y. Luo, J. D. Sanchez-Yamagishi, K. Watanabe,
T. Taniguchi, E. Kaxiras, R. C. Ashoori, and P. Jarillo-Herrero, Nature {\bf 556}, 80 (2018).

\bibitem{TBLG-2} Y. Cao, V. Fatemi, S. Fang, K. Watanabe, T. Taniguchi, E. Kaxiras, and P. Jarillo-Herrero, Nature {\bf 556}, 43 (2018).

\bibitem{organicDirac} M. Hirata, K. Ishikawa, G. Matsuno, A. Kobayashi, K. Miyagawa, M. Tamura, C. Berthier, and K. Kanoda, Science {\bf 358}, 1403 (2017).


\bibitem{TBLG-3} X. Lu, P. Stepanov, W. Yang, M. Xie, M. Ali Aamir, I. Das, C. Urgell, K. Watanabe, T. Taniguchi, G. Zhang, A. Bachtold, A. H. MacDonald, D. K. Efetov, arXiv:1903.06513

\bibitem{Kekule-MA-TBLG} For a possible realization of KVBS in MA-TBLG, see X. Y. Xu, K. T. Law, and P. A. Lee, Phys. Rev. B {\bf 98}, 121406 (2018) and Y. D. Liao, Z. Y. Meng, X. Y. Xu, arXiv:1901.11424.



\end{thebibliography}
\end{document}